%% file: apssamp.tex
\documentclass[%
 reprint,
 amsmath,amssymb,
 aps,
]{revtex4-2}

\usepackage{graphicx}% Include figure files
\usepackage{dcolumn}% Align table columns on decimal point
\usepackage{bm}% bold math
\usepackage{hyperref}% add hypertext capabilities
\renewcommand{\equationautorefname}{Eq.}
\makeatletter
\def\Hy@autoref@equation#1{%
  \begingroup
  \edef\@tempa{\endgroup
    \noexpand\hyperref[#1]{\equationautorefname~(\noexpand\ref*{#1})}}%
  \@tempa
}
\makeatother

\usepackage{siunitx}
\usepackage{physics}
\usepackage[english]{babel}
\renewcommand{\selectlanguage}[1]{}

\begin{document}

\preprint{APS/123-QED}

\title{High-yield fabrication of micromirror templates via feedback-controlled laser ablation}
\author{Daniel Allepuz-Requena}
\email{dalre@dtu.dk}
\author{Jonas S. Neergard-Nielsen}
% \email{alexander.huck@fysik.dtu.dk}
\author{Alexander Huck}
\email{alexander.huck@fysik.dtu.dk}
\author{Ulrik L. Andersen}
\email{ulrik.andersen@fysik.dtu.dk}
\affiliation{Center for Macroscopic Quantum States (bigQ), Department of Physics, Technical University of Denmark, Fysikvej, 2800 Kongens Lyngby, Denmark}

\date{\today}% It is always \today, today,

\begin{abstract}
We present a high-yield method for fabricating concave micromirror templates in silica using feedback-controlled CO$_2$ laser ablation with precise in situ positioning. Real-time monitoring of the white-light emission generated during ablation is used to terminate laser exposure, thereby reducing shot-to-shot variability in mirror depth and radius of curvature. To ensure reproducible single-shot processing across different substrates, the sample position relative to the laser focus is calibrated using an in situ phase-scanning interferometric microscope integrated into the fabrication workflow. The method enables reliable fabrication of shallow mirror templates with tunable radii of curvature spanning from approximately $\SI{20}{\micro\meter}$ to several hundred micrometers, with relative geometric variances as low as 3\%. The suitability of the fabricated mirrors for optical resonators is verified by realizing a compact plano-concave Fabry--Perot microcavity with a finesse of $3.7\times10^4$ at telecom wavelengths. The setup provides a simple and automated route to reproducible micromirror fabrication for applications in cavity quantum electrodynamics and cavity optomechanics.
\end{abstract}

\maketitle

%\tableofcontents

\input{content}
\bibliography{references, extra}% Produces the bibliography via BibTeX.

\end{document}

%% file: content.tex
\section{Introduction}
Building optical cavities with micrometer-scale lengths requires the use of shallow mirrors. In silicon substrates, low-loss concave micromirrors can be fabricated by etching~\cite{trupke_microfabricated_2005,ow_fabrication_2010, biedermann_ultrasmooth_2010,bao_concave_2017, fait_high_2021,jin_micro-fabricated_2022}. In silica substrates, which offer advantages due to their wider transparency window, ultra-smooth concave features can instead be realized by laser ablation~\cite{ muller_ultrahigh-finesse_2010,hunger_laser_2012}. In this approach, ablation is performed using a focused high-power CO\textsubscript{2} laser beam operating at a wavelength of approximately $\SI{10}{\micro\meter}$, which produces Gaussian-like surface depressions. Recent advances in laser ablation techniques have focused on mitigating the frequency splitting of polarization eigenmodes arising from mirror asymmetries~\cite{uphoff_frequency_2015, takahashi_novel_2014}, as well as on achieving smaller cavity mode volumes~\cite{greuter_small_2014,ruelle_optimized_2019}, which are particularly beneficial for cavity QED experiments~\cite{volz_measurement_2011, steiner_single_2013, albrecht_coupling_2013, kaupp_purcell-enhanced_2016, takahashi_cavity-induced_2017, gallego_strong_2018, takahashi_strong_2020, brekenfeld_quantum_2020, herrmann_coherent_2024}.\par

To a lesser extent, efforts have also been directed toward improving fabrication yield and reducing the variance in the resulting mirror geometries. A feedback-based method relying on real-time monitoring of white-light emission during ablation was shown to significantly reduce shot-to-shot variability~\cite{petrak_feedback-controlled_2011}. More recently, tight control of the micromirror profile has been demonstrated using a multistep approach that integrates ablation with in situ imaging~\cite{gao_profile_2025}. The single-shot nature of the real-time feedback approach enables access to smaller radii of curvature, and therefore smaller mode volumes, than the latter adaptive method.\par

High-yield and fabrication robustness are particularly important when working with specialized substrates, particularly when the cost of discarding a sample is high, or when fabricating arrays of nominally identical micromirrors~\cite{wachter_silicon_2019}. These considerations are especially relevant in cavity optomechanics. Mirror substrates engineered with phononic crystals have been the key to recent achievements in room-temperature quantum cavity optomechanics~\cite{saarinen_laser_2023,huang_room-temperature_2024-1, xia_motional_2025, allepuz-requena_mitigating_2026}. In this context, microcavities are attractive because the magnitude of the single-photon optomechanical coupling scales inversely with cavity length~\cite{aspelmeyerCavityOptomechanics2014}. Fiber-based microcavities that incorporate a membrane resonator benefit from their short length, but can suffer from excess noise arising from the thermomechanical motion of the mirrors~\cite{flowers-jacobsFibercavitybasedOptomechanicalDevice2012,rochauDynamicalBackactionUltrahighFinesse2021, saarinen_laser_2023, tenbrake_direct_2024}. This limitation can be mitigated by patterning the mirror substrate with a phononic crystal, which suppresses mechanical motion around a frequency of interest~\cite{saarinen_laser_2023,huang_room-temperature_2024-1, xia_motional_2025, allepuz-requena_mitigating_2026}.\par
In this work, we present a mirror-template fabrication setup that combines real-time feedback with precise sample positioning. An in situ phase-scanning interferometric microscope is used to reliably position the sample at the focal plane of the ablation laser, ensuring reproducible single-shot fabrication across different substrates. We demonstrate the fabrication of shallow mirror templates with adjustable radii of curvature ranging from tens of micrometers to the millimeter scale. Finally, we construct Fabry--Perot microcavities by coating the fabricated mirrors with low-loss distributed Bragg reflector coatings and verify the suitability of the templates for realizing high-finesse optical resonators.

\section{Methods}
\begin{figure}[h]
    \centering
    \includegraphics[width=0.9\linewidth]{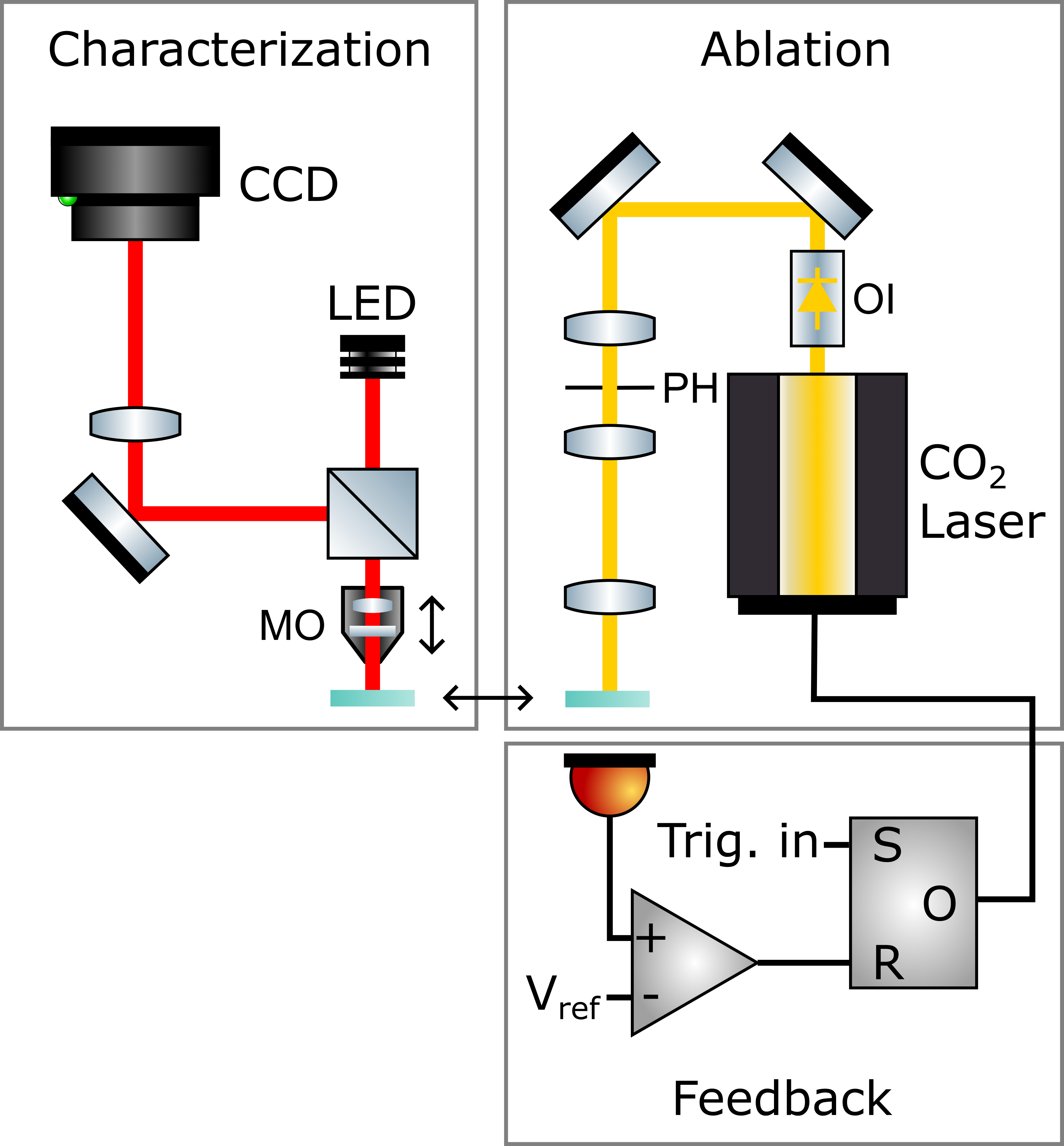}
    \caption{Schematic view of the mirror fabrication setup. The sample is moved using a motorized linear stage from the characterization side, a scanning-phase interferometric microscope, to the ablation part, where the beam of a CO\textsubscript{2} laser is focused on the sample to create a concave indentation. The sample is mounted on a motorized 3-axis translation stage (Physik Instrumente M112.1DG1) allowing for relative movement of the sample and alignment to the focus of the infrared laser beam. An additional motorized long travel range is used to move the 3-axis stage together with the sample from the ablation to the  characterization part and vice versa.
    MO: Mirau objective (movable along the optical axis using a piezoelectric stack), PH: pinhole, OI: optical isolator.}
    \label{fig:setup-drawing}
\end{figure}

A schematic view of our setup is shown in Fig.~\ref{fig:setup-drawing}. The setup consists of two parts, one part for laser ablation and another part for characterization. The sample, e.g. a substrate made of SiO$_2$ or a conventional optical fiber, can be moved between both parts using a motorized stage.

\subsection{Laser ablation with feedback}
 We use the focused beam of a $\SI{10.6}{\micro\meter}$ wavelength IR laser (Synrad v40 Firestar) to ablate the surface of the target sample. The laser is used at its nominal continuous wave optical power of $\SI{40}{\watt}$. An optical isolation stage placed behind the laser is used to prevent back-reflections from re-entering the laser cavity. To obtain a near Gaussian beam profile, we use a mode-cleaning telescope consisting of a $\SI{300}{\micro\meter}$ diameter gold-coated pinhole placed in the focal plane between two $f=\SI{50}{\milli\meter}$ plano-convex ZnSe lenses. Finally, the filtered beam is focused onto the sample using an additional plano-convex ZnSe lens, where the focal length partially determines the shape of the ablated features on the substrate.\par
 When the IR beam is absorbed by the sample, ablation occurs near the beam focus. White light is emitted during ablation, which is measured by a $\SI{13}{\milli\meter^2}$ active area biased Si photodiode (Thorlabs DET36A/M). Its signal forms the basis of the feedback system, once it reaches a predefined threshold, the IR laser stops and the ablation process finishes~\cite{petrak_feedback-controlled_2011}.\par
\begin{figure}[h]
    \centering
    \includegraphics[]{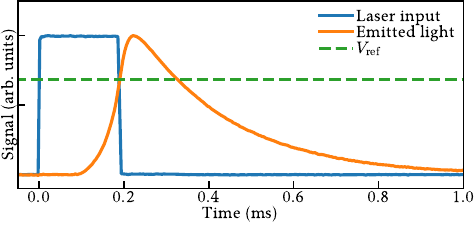}
    \caption{Signals recorded during the fabrication of a mirror template using feedback-controlled laser ablation. The blue curve is the TTL modulation signal sent to the laser. The orange curve is the white-light emission measured by the photodetector behind the target sample. The dashed green line represents the reference voltage used in the voltage comparator.}
    \label{fig:feedback-signals}
\end{figure}

We implement this protocol with an electronic circuit consisting of two main components: a voltage comparator (TI LM311P) and a Set-Reset latch (TI 74LS279). Ablation starts after a user-generated signal primes the latch to the high logic state, the output drives the laser into emission. The voltage comparator resets the latch to its low logic state once the photodetector signal exceeds a user-defined reference voltage $V_\text{ref}$. The laser emission stops and the ablation process is concluded. Time traces of the relevant signals are recorded by an oscilloscope, an example of which can be seen in Figure \ref{fig:feedback-signals}.\par
The simple implementation presented here is a cost-effective alternative to the system implemented in previous work which uses a Field-Programmable Gate Array (FPGA) device~\cite{petrak_feedback-controlled_2011}. The simplification is possible due to our laser implementing its own ``tickle" pulse generator, which are $\SI{1}{\micro\second}$ short pulses used to keep the gain medium ionized.\par

\subsection{Interferometric characterization}
After ablation, the geometry is characterized using a  home-built phase-scanning interferometry microscope. The key element of the interferometer is a Mirau microscope objective (Nikon CF IC EPI Plan DI 20×A) mounted on a linear translation stage (Thorlabs PT1/M) retrofitted with a piezoelectric chip stack (Thorlabs PC4QM). We use a light-emitting diode (LED) at $\SI{617}{\nano\meter}$ (Thorlabs M617L4) to create a magnified interferometric image that we record with a monochrome CCD camera (Thorlabs) using a lens $f=\SI{500}{\milli\meter}$. The height profile of the sample can be reconstructed by capturing a series of images at different objective-to-sample distances, which we control using the piezoelectric actuator mounted in the objective base. The objective is scanned approximately $\SI{8}{\micro\meter}$ along the optical axis. The relative phase between two pixels can be extracted from the evolution of their values. Then a two-dimensional phase-unwrapping algorithm \cite{herraez_fast_2002} is used to recover the height profile of the sample.

\subsection{Fabrication procedure}
\begin{figure}
    \centering
    \includegraphics[width=\linewidth]{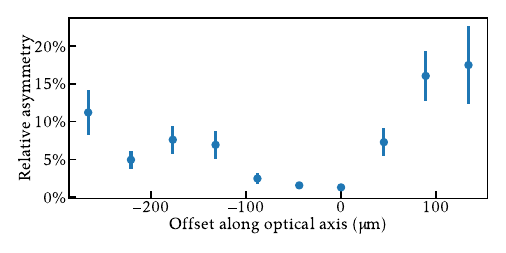}
    \caption{Average relative asymmetry of the mirror template's curvature as it is offset along the optical path of the ablating laser. The error bars represent the standard deviation of the asymmetry of 10 templates fabricated at that offset.}
    \label{fig:offset_calibration}
\end{figure}

The procedure used to fabricate a mirror template is as follows. The apparatus is controlled through a Graphical User Interface (GUI) based on the Qudi software suite~\cite{binder_qudi_2017}. The program communicates with a microcontroller responsible for generating the logic signal that primes the feedback system and for setting the reference voltage used by the comparator. The fabrication of a single feature proceeds as follows:

\begin{enumerate}
    \item The sample is initially positioned on the characterization side. It is brought to the focal plane of the interferometric microscope using an autofocus procedure. Owing to the short coherence length of the LED ($\SI{8}{\micro\meter}$, according to the manufacturer’s specifications), interference fringes are observed only when the microscope is focused close to the sample surface, providing a well-defined reference position. A live view of the interferometric image also allows the user to precisely select the ablation location, for example, the core of an optical fiber.
    \item The sample is translated to the ablation side by a calibrated displacement in all three spatial directions, such that the center of the interferometric field of view coincides with the focal point of the IR laser. This calibration was performed by ablating at different axial offsets relative to the IR laser focus and selecting the offset that minimized feature asymmetry. Ten offsets are tested in a region of $\SI{400}{\micro\meter}$ along the optical axis. Ablation is performed 10 times per offset and the average relative asymmetry of their curvature is measured (see Results section for definition). Fig.~\ref{fig:offset_calibration} shows how the asymmetry is minimized within a region that we associate with the focal plane.
    \item The feedback-controlled ablation process is initiated by the microcontroller.
    \item The sample is returned to the characterization side, where a height map of the fabricated feature is acquired.
\end{enumerate}
Our setup is completely automated and can create and characterize mirrors in the same target substrate at an approximate rate of 1 mirror per minute.
\section{Results}
\begin{figure}
    \centering
    \includegraphics[width=\linewidth]{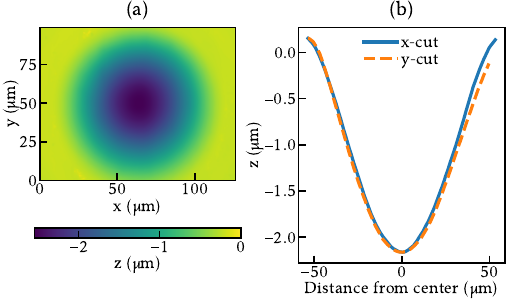}
    \caption{(a) Height map of a fabricated feature obtained through phase-scanning interferometry. (b) Height profiles along the horizontal and vertical cuts that intercept the center of the feature.}
    \label{fig:heightmap}
\end{figure}

In this section, we discuss the mirror geometries that have been achieved, as well as the performance of the feedback system. The geometry of the mirror templates is measured using the phase-scanning interferometric microscope, which is able to reconstruct height profiles with a horizontal spatial resolution of roughly $\SI{100}{\nano\meter}\mathrm{/pixel}$. The scaling factors have been calibrated by measuring the profiles of atomic force microscope calibration gratings. An example of a height map obtained by phase-scanning interferometry is shown in \autoref{fig:heightmap}. The depth $d$ of the feature (height at its deepest point) can be measured directly from the height map. We characterize the curvature using the same parameters that appear in Ref.~\cite{ruelle_optimized_2019}. Two curvature radii $R_a$ and $R_b$ are defined to characterize the possible asymmetry of the feature. They represent the radius of curvature along the major and minor axes of the elliptic paraboloid used to approximate the lower part of the mirror. $R_a$ and $R_b$ are obtained by fitting the center of the height map with the following expression:
\begin{equation}
    \begin{aligned}
        z(x,y) \propto &\frac{1}{2 R_a} (x \cos{\phi} - y\sin{\phi})^2 + \\ &+ \frac{1}{2 R_b} (x \sin{\phi} + y\cos{\phi})^2,
\end{aligned}
\end{equation}

where $\phi$ is the angle formed by the major axis of the ellipse and the horizontal axis of the image. In the following discussions, we use the mean value $\left( R_a + R_b \right)/2$ as a measure of the radius of curvature, while the relative asymmetry of the mirror is quantified by ${\lvert R_a - R_b \rvert}/\qty(R_a + R_b)$. 

\subsection{Repeatability and yield}
\begin{figure}
    \centering
 
    \includegraphics[width=\linewidth]{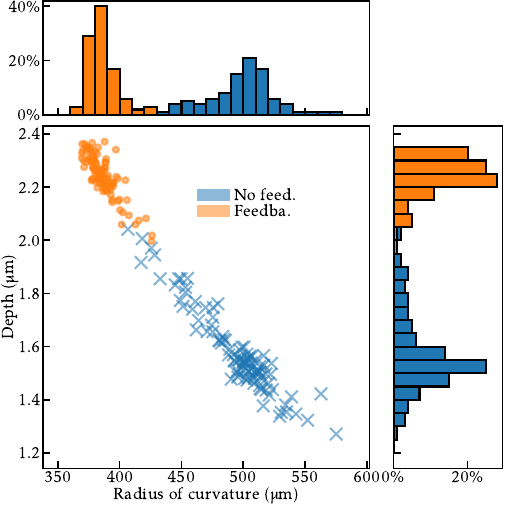}
    \caption{Comparison of the geometry distribution of 100 mirrors without (blue crosses) and with feedback (orange circles). In the absence of feedback, the modulation signal to the laser is a square pulse with a constant duration of $\SI{184}{\micro\second}$. 
    }
    \label{fig:feedback_comparision}
\end{figure}

We compare 100 mirror templates created with and without feedback. All mirror templates have been ablated on the same fused silica substrate on which the IR laser has been focused using a $f=\SI{100}{\milli\meter}$ lens. \autoref{fig:feedback_comparision} shows the distribution of depth and radius of curvature of the mirrors with and without feedback. The feedback system clearly reduces the variation of the mirror shape. The relative variance, calculated as the standard deviation to mean value ratio, of the mirror depth is reduced from 10\% to 3\%. The relative variance of the radius of curvature is reduced from 6\% to 3\%. The average feature asymmetry is around $3\%$ for both cases. The improvement is due to the feedback system correcting for the unpredictable delay between the control pulse sent to the laser and the start of the ablation. Without feedback, this delay introduces a variance in the active ablation time, which ultimately translates into changes in mirror geometry. \par
\begin{figure}
    \centering
    \includegraphics[width=\linewidth]{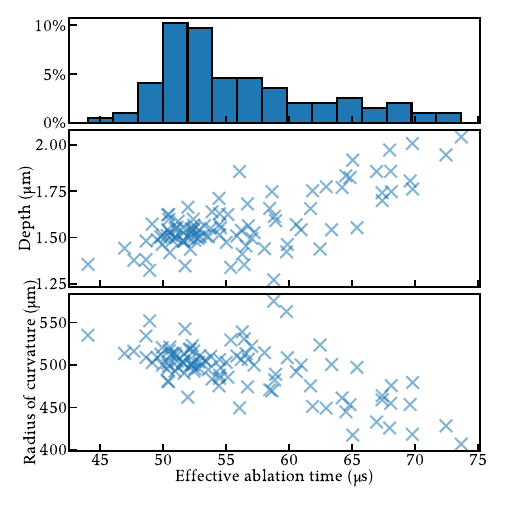}
    \caption{Without feedback, the random delay of the start of ablation sets the variance of the effective ablation time. Top:  histogram of the effective ablation time. Middle and bottom: depth and radius of curvature as a function of effective ablation time.}
    \label{fig:ablation_time}
\end{figure}

We have investigated the effects of ablation time by measuring the time from the ablation start (defined as the signal surpassing 10\% of its maximum value) to the falling-edge of the laser excitation signal. Without feedback, the effective ablation time has a relative variance of 13\%. As can be seen in Fig.~\ref{fig:ablation_time}, its distribution is non-Gaussian and positively skewed. As also seen in the figure, both the depth and radius of curvature are strongly correlated with the effective ablation time. With feedback, the variance of the effective ablation time is reduced to 8\%, which is the cause of the reduction of geometry variance seen in Fig.~\ref{fig:feedback_comparision}.\par
The delay of the ablation start might be caused by the laser responding to the signal with different rise times depending on the state of the gain medium, which will depend on the temporal proximity to the tickle pulses generated by the laser internal circuitry ($\SI{1}{\micro\second}$ in length at a $\SI{1}{\kilo\hertz}$ repetition rate). In addition, the delay in ablation might also be caused by debris on the surface of the target substrate. Ultimately, we expect the feedback system to be limited by power fluctuations of the laser, which according to the manufacturer are $\pm 3\%$.\par


\begin{figure*}
    \centering
    \includegraphics[width=17cm]{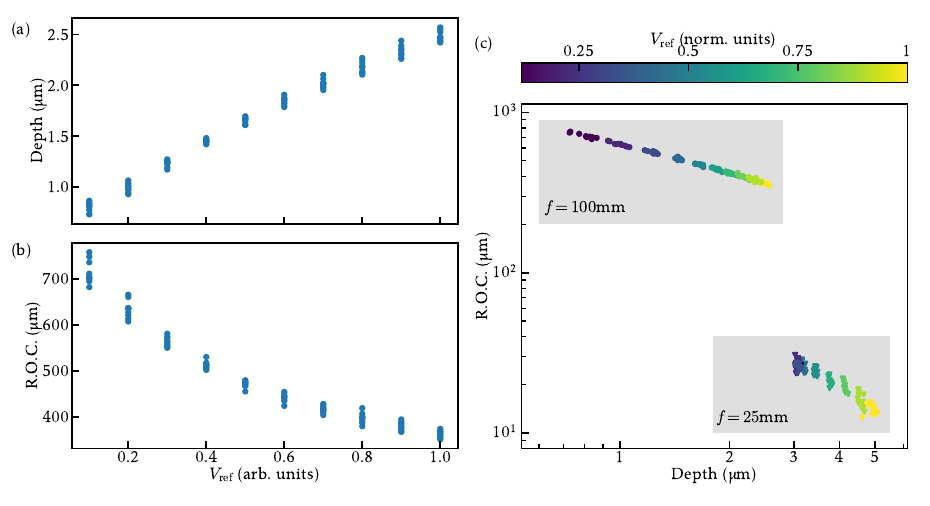}

    \caption{Reference voltage $V_\mathrm{ref}$ dependence of the mirrors' (a) depths and (b) radii of curvature. (c) Scatter plot of the radius of curvature and depth of each mirror template. We include data from 100 mirrors fabricated using a $\SI{25}{\milli\meter}$ lens. Ten features were fabricated at each reference voltage. $V_\mathrm{ref}$ is normalized to the maximum in the units of the digital-to-analog converter used.}
    \label{fig:vrefs}
\end{figure*}

\subsection{Range of geometries}
 The $\SI{100}{\milli\meter}$ focal length has been selected because it results in mirror templates with radii of curvature around $\SI{300}{\micro\meter}$ at a mirror depth of around $\SI{2}{\micro\meter}$. These mirrors are to be used in a cavity optomechanics experiment and thus we do not aim for the smallest radius of curvature. Nevertheless, using a shorter focal length allows us to create templates with smaller radius of curvature, which can be used in low-volume optical cavities more suitable for cavity QED experiments. Using a $\SI{25}{\milli\meter}$ focal length lens, we can reliably fabricate mirrors with a radius of curvature of about $\SI{20}{\micro\meter}$.\par

The reference voltage triggering the laser to stop ($V_\text{ref}$) can be tuned to obtain different mirror geometries. \autoref{fig:vrefs} (a) and (b) show the wide range of mirror geometries that can be obtained by adjusting the value of $V_\text{ref}$. Figure \ref{fig:vrefs} (c) shows the verification of the power law relating the radius of curvature and depth found in previous work~\cite{hungerFiberFabryPerot2010,ruelle_optimized_2019}. We also include geometries obtained when using a $f=\SI{25}{\milli\meter}$ lens to showcase the wide range of geometries accessible using our setup. An even wider range could be accessed by adjusting the power of the IR laser (i.e. by means of a polarizer) or offsetting the sample from the laser's focus.

\subsection{Realization of an optical microcavity}

\begin{figure}
    \centering
    \includegraphics[width=0.9\linewidth]{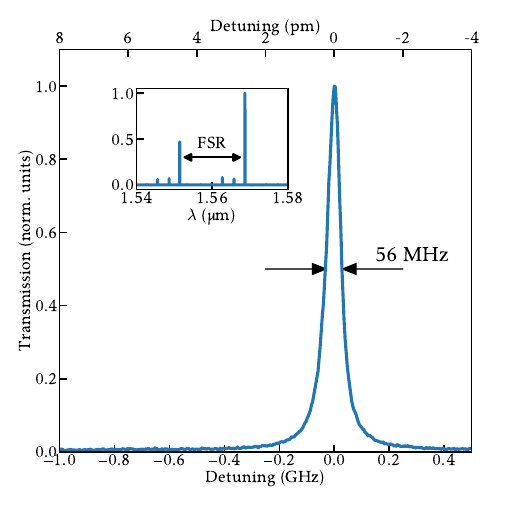}        

    \caption{Transmission spectrum of the fabricated microcavity obtained by scanning the laser frequency across resonance and measuring the transmitted signal. The frequency axis has been calibrated by generating phase-modulation sidebands at a known frequency. The inset figure shows a wide scan of the laser wavelength $\lambda$ across two resonances.}
    \label{fig:transmission}
\end{figure}

We realize a compact high-finesse plano-concave microcavity to test the adequacy of the mirrors. The concave mirror is realized by fabricating a $R\approx\SI{300}{\micro\meter}$ concave mirror on one side of a $\SI{500}{\micro\meter}$ thick fused-silica square substrate. The planar mirror consists of a $\SI{1.1}{\milli\meter}$ thick, double-sided polished, square silicon substrate. Both substrates are coated with low-loss, low-transmission distributed Bragg reflector (DBR) using ion beam sputtering, and an anti-reflective coating is deposited on the other side. The DBRs are made of alternating layers of SiO$_2$/Ta$_2$O$_5$ with target transmissions of 10\,ppm (planar mirror) and 100\,ppm (curved mirror) for $\SI{1550}{\nano\meter}$ light. The cavity is built by stacking both mirrors around a spacer and clamping the stack with copper clamps. The spacer consists of single-sided Kapton tape with a thickness of roughly $\SI{65}{\micro\meter}$ (specified by the manufacturer). Thus, the cavity length is fixed and not tunable.

We use a widely-tunable laser (Toptica CTL 1550) to characterize the optical properties of the micro-cavity. We measure a free spectral range (FSR) of $\SI{2.119(1)}{\tera\hertz}$ by doing a continuous scan of the laser wavelength from $\SI{1510}{\nano\meter}$ to $\SI{1630}{\nano\meter}$ and recording the transmission with an InGaAs photodiode (Fig.~\ref{fig:transmission} inset). We infer a cavity length of $\SI{70.75}{\micro\meter}$ and a mirror curvature radius of approximately $\SI{270}{\micro\meter}$. The linewidth of the optical resonance is measured by fitting the transmission signal with a Lorentzian peak. The laser frequency is scanned around the resonance frequency at a rate of  $\SI{40}{\giga\hertz/\second}$. To calibrate the transmission spectrum, we modulate the phase of the beam with a sine at a frequency of $\SI{500}{\mega\hertz}$ to create two additional resonant transmission peaks around the optical mode. The linewidth is then measured by fitting a Lorentzian peak to the frequency-calibrated transmission spectrum. Fig.~\ref{fig:transmission} shows the resonance peak with the minimum observed linewidth of $\SI{56}{\mega\hertz}$, corresponding to a finesse of around $37000$ at a wavelength of $\SI{1568}{\nano\meter}$. The finesse is limited due to the monolithic cavity construction as we cannot set the cavity resonant condition to the wavelength that maximizes reflection by the DBRs. The birefringence of the cavity was tested by preparing the input light in a circularly polarized state to no perceivable effect on the transmission spectrum. This indicates that, at this length and finesse, the birefringence is significantly smaller than the linewidth.

\section{Conclusions}
We have demonstrated a robust and automated platform for the fabrication of low-roughness concave micromirror templates in silica using feedback-controlled CO$_2$ laser ablation. By combining real-time monitoring of ablation-induced white-light emission with precise sample positioning relative to the laser focus, the system achieves highly reproducible single-shot fabrication across a wide range of geometries. Compared to open-loop ablation, the implemented feedback scheme substantially reduces shot-to-shot variations in both mirror depth and radius of curvature, yielding relative spreads of approximately 3\%. Through tuning of the feedback threshold and lens focusing power, our setup enables the fabrication of micromirrors with radii of curvature spanning from the tens of micrometers to the millimeter scale. We validated the suitability of the templates by realizing a compact monolithic Fabry–Perot microcavity that exhibits a finesse of up to $3.7\times10^4$ at telecom wavelengths. Despite residual asymmetry in the mirror geometry, we do not observe mode splitting in the fundamental optical mode.\par
Our system enables reliable one-shot fabrication of micromirrors on valuable or pre-processed substrates, where repeat ablation is not an option. In its current implementation, the setup has been used to fabricate vibrationally isolated concave micromirrors at the center of phononic crystals etched on silica substrates~\cite{allepuz-requena_mitigating_2026}. More generally, the platform enables fabrication of large arrays of open-access microcavities in silica with performance comparable to state-of-the-art arrays realized in silicon substrates~\cite{wachter_silicon_2019}. Beyond regular arrays, the ability to define arbitrary micromirror layouts could enable vertical Fabry–Perot cavities integrated directly above planar photonic circuits~\cite{cheng_harnessing_2025}.  

\section*{Acknowledgments}
We acknowledge funding support from The Danish Council for Independent Research | Natural Sciences (grant no. 11-108077) and the Danish National Research Foundation (bigQ, DNRF0142). The authors thank Ilya Radko for preliminary work on the experimental setup. The authors also thank Guillem Allepuz Requena for help with hardware implementation.

\section*{Data availability}
A detailed description of the electronics implementing the feedback system, the designs for a printed circuit board with grounded casing, and the software that controls the experiment are freely available online~\cite{repository}. The data that supports the findings of this study are available from the corresponding author upon reasonable request.